\documentstyle[preprint,aps,version2]{revtex} 
\def\PL #1 #2 #3 {Phys. Lett.~{\bf#1} (#2) #3}
\def\NP #1 #2 #3 {Nucl. Phys.~{\bf#1} (#2) #3}
\def\ZP #1 #2 #3 {Z.~Phys.~{\bf#1} (#2) #3}
\def\PR #1 #2 #3 {Phys. Rev.~{\bf#1} (#2) #3}
\def\PRD #1 #2 #3 {Phys. Rev.~D {\bf#1} (#2) #3}
\def\PP #1 #2 #3 {Phys. Rep.~{\bf#1} (#2) #3}
\def\PRL #1 #2 #3 {Phys. Rev.~Lett.~{\bf#1} (#2) #3}

\def\ie {{\it i.e}.}
\begin{document}

\hfill\parbox{2in}{\baselineskip14.5pt
{\bf MAD/PH/814}\\
June 1994}

\vspace{.25in}

\begin{title} Evidence for a Hard Pomeron in Perturbative QCD
\end{title}
\author{Hassan~N.~Chehime and Dieter~Zeppenfeld}
\begin{instit}
Department of Physics, University of Wisconsin, Madison, WI 53706, USA
\end{instit}

\thispagestyle{empty}

\nonum\section{abstract}
\baselineskip14.5pt
The $t$-channel exchange of two gluons in a color singlet state represents
the lowest order approximation to the Pomeron. This exchange mechanism is
thought to also explain the formation of rapidity gaps in dijet events at
the Tevatron. At the perturbative level this requires suppressed gluon
emission in the rapidity interval between widely separated jets,
analogous to color coherence effects in $t$-channel photon exchange.
By calculating the imaginary part of the two gluon, color singlet exchange
amplitude we show how this pattern does emerge for gluon emission at small
transverse momenta. At large $p_T$ the radiation pattern characteristic for
color octet single gluon exchange is reproduced.

\newpage

\section{Introduction}

Over the past three decades the Pomeron model has been very successful in
describing a wealth of data on hadronic collisions: the total cross
section, elastic scattering, and single diffractive
scattering~\cite{Grib,low,lipatov,lan}. As viewed from
perturbative QCD the Pomeron may be thought of as the $t$-channel exchange
of two gluons in a color singlet state~\cite{low}. This simple picture, to be
called the Low-Nussinov Pomeron in the following, has been refined in
subsequent years in terms of regeizzed gluon ladders leading to the Lipatov
hard Pomeron~\cite{lipatov}.

One of the defining features of diffractive scattering is the pattern of final
state hadrons. In a collider configuration hadronization occurs in narrow cones
around the beam directions while the central rapidity region is free of
produced hadrons (rapidity gap). The same phenomenon has recently been observed
in {\it hard} scattering events at the Tevatron~\cite{brandt}. The D0
Collaboration has studied a sample of dijet events with jet transverse energies
in excess of 30 GeV. In about $ 0.5$\%  of all events with widely
separated jets no sign of hadronic activity is observed  between the two jets.

This observation is consistent with predictions for rapidity gap event
rates within the Pomeron model~\cite{bjgap}. Bjorken estimated the size
of the Low-Nussinov Pomeron exchange contribution and found that a fraction
$f\approx 10$\% of all dijet events should be due to this mechanism. The other
factor in his estimate is the ``survival probability'' $P_s$, \ie\ the
probability that the underlying event does not fill the rapidity gap which is
present at the parton level. The survival probability is expected to be of
order 10\% at the Tevatron, with large uncertainties~\cite{bjgap,gotsman},
leading to a rapidity gap fraction $fP_s=0.01$, as compared to the observation
at the 0.005 level.

Both Pomeron and $t$-channel photon exchange are believed to lead to rapidity
gaps since both scattering processes proceed via the exchange of a $t$-channel
color singlet object. In the case of $\gamma$ exchange in forward
$qq \rightarrow qq$ scattering, color coherence~\cite{colcoh} between initial
and final state gluon radiation is known to lead to an exponentially
suppressed gluon emission probability into the rapidity region between the two
final state quarks~\cite{fletcher}. After
hadronization this translates into suppressed hadronic activity between the
two quark jets, thus allowing the formation of a rapidity gap. Due to its
color singlet structure one expects that the same would occur for the
$t$-channel exchange of a Pomeron. However, since the Low-Nussinov Pomeron is
an extended object with colored constituents, the validity of this analogy
is not obvious: gluon radiation may resolve the internal color structure of
the Pomeron.

In this paper we investigate the conditions (if any) under which gluon
emission between the two final state quarks is suppressed in the Pomeron
exchange process . This radiation pattern is necessary for the interpretation
of
the sizable rate of rapidity gap events observed by D0 since electroweak
processes alone, {\it i.e.} $t$-channel $\gamma$, $Z$ or $W$ exchange, have
too small a cross section to account for the observed rate~\cite{chehime}.
We calculate the gluon radiation pattern for
the imaginary part of the $t$-channel exchange of two gluons in quark-quark
scattering. There are practical reasons why we do not attempt a full
calculation. The lowest order cross section for gluon radiation in two gluon
color singlet exchange is of order $\alpha_s^5$ and hence the full calculation
corresponds to a two-loop evaluation of 3-jet production at hadron colliders,
which is clearly beyond the scope of the present work. Instead, we aim at
a qualitative understanding of the radiation pattern only. Since the
imaginary part of the amplitude is known to dominate over the real part
in forward, elastic quark-quark
scattering via Pomeron exchange~\cite{cudell}, the imaginary part of the
emission process should be sufficient for our purposes.

The gluon radiation pattern for color singlet two-gluon exchange is then
compared to the ones obtained for (color
singlet) $\gamma$ exchange and for (color octet) gluon exchange.
For the exchange of a Low-Nussinov Pomeron we find that
the angular distribution of a single radiated gluon can be described in
terms of three components: a QED-like contribution which is consistent with
the formation of rapidity gaps, a QCD like term with gluon emission in the
backwards direction, and a collinear term which describes $q\to g$ splitting
and subsequent Pomeron exchange between the gluon and the second quark. For
the emission of gluons at low transverse momentum the QED-like component
is much larger than the QCD-like one. Since these soft gluons dominate the
overall gluon radiation, hadronization will produce
a pattern similar to the one expected for $t$-channel photon exchange.
Therefore Pomeron exchange should indeed lead to the
formation of rapidity gaps in hard scattering events. For those willing to
extrapolate to low $|t|$ this finding gives further credence to a QCD
explanation also of the soft Pomeron.

The main part of this paper is organized as follows. In Section~II we discuss
the most general color structure of the $qQ\rightarrow qQg$ process. Two of the
terms in the color decomposition define what we mean by ``color singlet
exchange in the $t$-channel''. We then apply this decomposition to the tree
level photon and single gluon exchange processes, which are later needed for
comparison, and discuss their radiation patterns. Section~III is technical and
describes our calculation of the imaginary part of  $qQ\rightarrow qQg$
scattering via two gluon exchange.  The resulting radiation pattern is then
discussed in Section~IV and Section~V gives our conclusions.

\section{Color Structure and Tree Level Amplitudes}

We start by analyzing the color structure and the resulting gluon radiation
patterns for the scattering of two quarks of different flavor at tree level.
The amplitudes for single gluon and single photon exchange already exhibit the
full color structure of the process $qQ\to qQg$ and these results will be
needed later for comparison with Pomeron exchange. We thus study the process
\begin{equation} \label{process}
q(p_1,i_1)\,Q(p_3,i_3) \longrightarrow q(p_2,i_2)\,Q(p_4,i_4)\,g(k,a)\; ,
\end{equation}
where $p_n$ and $k$ denote the quark and gluon momenta and their colors are
labeled by $i_n$ and $a$. Scattering involving anti-quarks is related by
crossing and we do not consider processes with initial state gluons.

In general, the amplitude of the $qQ\to qQg$ process can be written in terms
of four orthogonal color tensors, which we denote by
$O_{1}$, $O_{2}$, $S_{12}$, and $S_{34}$,
\begin{equation}\label{amp}
M = O_{1}M_{1} + O_{2}M_{2} + S_{12}M_{12} + S_{34}M_{34}\; .
\end{equation}
In an $SU(N)$ gauge theory they are given explicitly by
\begin{eqnarray}\label{colortensor}
S_{12} & = & {\frac {\lambda_{i_4i_3}^{a}} {2} }\; {\delta_{i_2i_1}} \\
S_{34} & = & {\frac {\lambda_{i_2i_1}^{a}} {2} }\; {\delta_{i_4i_3}} \\
O_{1} & = & \frac{-2}{N}\; ( S_{12} + S_{34} ) +
            {\frac {\lambda_{i_2i_3}^{a}} {2} }\; {\delta_{i_4i_1}} +
            {\frac {\lambda_{i_4i_1}^{a}} {2} }\; {\delta_{i_2i_3}} \\
O_{2} & = & {\frac {\lambda_{i_4i_1}^{a}} {2} }\; {\delta_{i_2i_3}} -
            {\frac {\lambda_{i_2i_3}^{a}} {2} }\; {\delta_{i_4i_1}}  \; .
\end{eqnarray}
$S_{12}$ marks the process where the quark $q$ keeps its color. Similarly,
$S_{34}$ multiplies the amplitude for $t$-channel color singlet exchange as
viewed from quark $Q$.  Within QCD, $O_{1}$ and  $O_{2}$ correspond to
$t$-channel color octet exchange as viewed from either of the two scattering
quarks.

Let us first apply this color decomposition to
the tree level QED and QCD amplitudes. The five Feynman graphs for
the QCD process are shown in Fig.~\ref{figone}. Lumping the momentum and
helicity dependence of the individual Feynman diagrams into reduced
amplitudes $A^{(1)}\cdots A^{(5)}$, one obtains for the QCD amplitudes
at tree level
\begin{eqnarray}\label{ampQCD}
M^{\rm QCD}_{12} & = &  \frac{-g^3}{2N}\;
              \left( A^{(1)} + A^{(2)} \right)
               \equiv  \frac{-g^3}{2N}\; A^{(12)}\\
M^{\rm QCD}_{34} & = &  \frac{-g^3}{2N}\;
              \left( A^{(3)} + A^{(4)}\right)
               \equiv \frac{-g^3}{2N}\; A^{(34)}\\
M^{\rm QCD}_{1}  & = &  \frac{-g^3}{4}\;
              \left( A^{(1)} + A^{(2)} + A^{(3)} + A^{(4)}\right) \\
M^{\rm QCD}_{2}  & = &  \frac{-g^3}{4}\;
              \left(A^{(1)} - A^{(2)} + A^{(3)} - A^{(4)} + 2A^{(5)}\right)
               \equiv \frac{-g^3}{4}\; A^{({\rm na})} \;.
\end{eqnarray}
The non-abelian three-gluon-vertex only contributes to the color octet
exchange amplitude $M_2$. Both $M_2$ and $M_1$ vanish for $t$-channel
photon exchange, while the color singlet exchange amplitudes are given by
\begin{eqnarray}\label{ampQED}
M^{\rm QED}_{12} & = &  -ge_qe_Q\;
              \left( A^{(3)} + A^{(4)} \right)
                   = -ge_qe_Q\; A^{(34)}\\
M^{\rm QED}_{34} & = &  -ge_qe_Q\;
              \left( A^{(1)} + A^{(2)}\right)
                   = -ge_qe_Q\; A^{(12)}\;,
\end{eqnarray}
where $e_q$ and $e_Q$ denote the electric charges of the two quarks.

The resulting color and polarization summed squared amplitudes,
\begin{equation}\label{sumM2}
\sum |M|^2 = {N^2-1 \over 2}N\, \sum_{\rm polarizations}
\left( |M_{12}|^2 + |M_{34}|^2
+ 2{N^2-4\over N^2}|M_1|^2 + 2|M_2|^2 \right)\; ,
\end{equation}
are shown in Fig.~\ref{QEDQCDtree}(a) for both the full QCD and the QED case
and for the corresponding color singlet contributions. Shown is the
dependence of $\sum |M^2|/s$ on the rapidity of the emitted gluon when all
other phase space parameters are kept fixed, namely the two quarks are
held at $p_T=30$~GeV and pseudorapidities $\eta_q=3$ and $\eta_Q=-3$ and
the gluon transverse momentum is chosen to be $p_{Tg}=2$~GeV. The details
of this choice are irrelevant: forward
scattering of the two quarks is a sufficient condition to obtain the
qualitative radiation pattern of Fig.~\ref{QEDQCDtree}.
In the QCD case the color octet contributions, via  the non-abelian
amplitude $A^{({\rm na})}$, lead to enhanced gluon emission
in the angular region between the two jets. For $t$-channel photon exchange
this region is essentially free of gluons due to color coherence
between initial and final state gluon radiation~\cite{colcoh,fletcher}:
gluon emission into the central region is exponentially suppressed as the
rapidity distance from the quarks increases. It is exactly this difference
between the QED and the QCD case which lets us expect the formation of
rapidity gaps for color singlet exchange in the $t$-channel.

Single gluon exchange also produces a color singlet exchange piece which is
defined by the $|M_{12}|^2+|M_{34}|^2$ term in Eq.~(\ref{sumM2}). Since its
shape is identical to the one for tree level photon exchange (dashed and
dash-dotted lines in Fig.~\ref{QEDQCDtree}(a)), one might be misled to
conclude that the color singlet contribution to single gluon exchange also
produces rapidity gaps. However, there is an important difference between
the two: the role of the amplitudes $A^{(12)}$ and $A^{(34)}$ is interchanged
in the QCD {\it vs.} QED color singlet exchange amplitudes.

The appearance of $A^{(12)}=A^{(1)} + A^{(2)}$ in the QED amplitude
$M^{\rm QED}_{34}$ corresponds to the first two Feynman graphs in
Fig.~\ref{figone}, {\it i.e.} emission of the final state gluon from the upper
quark line. Color coherence between these two graphs strongly suppresses gluon
emission except in the angular region between the initial and final direction
of the upper quark $q$ (dashed line in  Fig.~\ref{QEDQCDtree}(b)).
In forward quark
scattering via photon exchange the color $i_1$ of the initial quark $q$ is
thus transferred to a low mass color triplet object which emerges close to
the original $q$ direction. At lowest order this is the final state $q$, at
${\cal O}(\alpha_s)$ it is the $qg$ system. The situation is thus stable
against
gluon emission at even higher order for the QED case and gluon radiation is
suppressed in the rapidity range between the two final state quarks.

Formally, the QCD amplitude $M^{\rm QCD}_{34}$ also describes $t$-channel
color singlet exchange as seen from the lower quark line.
Proportionality to $A^{(34)}$ implies that the gluon is emitted from the
lower quark, $Q$, and hence preferentially between the initial $Q$-beam and the
final $Q$ directions (dash-dotted line in Fig.~\ref{QEDQCDtree}(b)).
Thus the color triplet $qg$ system, into which the
initial quark $q$ evolves, consists of a widely separated quark and gluon.
Higher order corrections will lead to strong gluon radiation into the angular
region between the two and thus also into the rapidity range between the
two final state quarks.

One thus finds that a color singlet exchange structure as identified by the
color tensors is not sufficient to infer the production of a rapidity gap in
the gluon radiation pattern. Stability of the pattern against multiple gluon
emission is another necessary requirement.
The shapes produced by $|M^{\rm QED}_{34}|^2$ and
$|M^{\rm QCD}_{34}|^2$ may now be taken as a gauge for the radiation pattern
produced by $t$-channel two-gluon exchange when the color index on the
$Q$-quark line is preserved. Only a QED like pattern can be expected to lead
to the formation of a rapidity gap in dijet events.

\section{The Imaginary Part of $M_{34}$}

A complete calculation of gluon emission in $qQ$ scattering via two-gluon
exchange involves squaring box-diagrams like the ones in Fig.~\ref{figthree}
and thus corresponds to a two-loop calculation of the 3-jet cross section
at hadron colliders, which is clearly beyond the scope of the present work.
Since $qQ\to qQ$ scattering via color singlet two gluon exchange is dominated
by its imaginary part at $|t|<<s$~\cite{cudell}, we restrict our attention to
the imaginary part of $M_{34}$, as defined by unitarity,
\begin{eqnarray}
2(Im\,T)_{fi} & \equiv & \frac{1}{i}(T_{fi} - {T^{\dagger}}_{fi})
\nonumber \\
& = &  \int \prod_{j=1}^{n} \frac{d^{3}p_{j}}{(2\pi)^{3}2p^{0}_{j}}
   (2\pi)^{4}\delta^{4}(\sum_{j=1}^{n}p_{j} - p_{i}) {T^{\dagger}}_{fn}T_{ni}
\; .    \label{unitarity}
\end{eqnarray}
Here $n$ denotes the number of on-shell intermediate particles and $i=|qQ>$
and $f=|qQg>$ symbolize the initial and final state.

To lowest order, ${\cal O}(g^5)$, only $n=2$ and $n=3$ particle intermediate
states contribute to $Im(M_{34})$. Representative Feynman graphs are depicted
in Fig.~\ref{figthree}. Row a) shows graphs corresponding to $n=2$ particle
intermediate states $|qQ>$. Factoring out color factors and coupling constants,
the contributions to ${T^{\dagger}}_{fn}T_{ni}$ are given by the products of
the tree level amplitudes listed below the individual Feynman diagrams. Here
$B_i$ corresponds to the 1-gluon exchange graph in $qQ$ elastic scattering
and the $A_f^{(j)}$ are the tree level $qQ\to qQg$ amplitudes of
Fig.~\ref{figone}. When considering $qQg$ 3-particle intermediate states
at order $g^5$ the $3\to 3$ amplitude decomposes into one disconnected
particle and a tree level $2\to 2$ amplitude. The three possibilities are
shown in rows b), (disconnected gluon), c) (disconnected quark $Q$), and d)
(disconnected quark $q$) of Fig.~\ref{figthree}. $B_f$ refers to the tree level
$qQ\to qQ$ 1-gluon exchange graph and $C_f^{(j)}$ and $D_f^{(j)}$ ($j=1,2,3$)
correspond to the three tree level graphs contributing to $qg\to qg$ and
$Qg\to Qg$ scattering, respectively.

Because of the disconnected particle in the $3\to 3$ amplitudes the phase
space integral in the unitarity relation of Eq.~(\ref{unitarity}) reduces to
a solid angle integral over the direction of one of the two intermediate
state particles participating in the scattering, taken in the two particle
rest frame. Using the abbreviations $A^{(12)}$, $A^{(34)}$, and $A^{(na)}$
(see Eq.~(\ref{ampQCD})--(10)) for the gauge invariant combinations of tree
level $qQ\to qQg$ amplitudes, one finds
\begin{eqnarray}
Im(M_{34}) = \frac{g^5}{16N^2}\frac{1}{32\pi^2}\int d\Omega  \biggl(
&-&B_i\, \left[-N^2A_f^{(na)}+(N^2-2)A_f^{(12)}+(N^2-4)A_f^{(34)} \right]
\nonumber \\
&-&\left[ N^2A_i^{(na)}+(N^2-2)A_i^{(12)}+(N^2-4)A_i^{(34)} \right]\, B_f
\nonumber \\
&+& 2A_i^{(34)}\,\left[ (N^2-1)C_f^{(1)}-C_f^{(2)}+N^2C_f^{(3)} \right]
\nonumber \\
&+& \left[ (N^2-4) A_i^{(12)}+(N^2-2)A_i^{(34)} \right]
    \left[ D_f^{(1)}+D_f^{(2)} \right]
\nonumber \\
&+&N^2A_i^{(na)}\left[ D_f^{(1)}-D_f^{(2)}+2D_f^{(3)}\right] \biggr) \; .
\label{M34tot}
\end{eqnarray}

Because of the massless $t$-channel gluon propagators in the Feynman graphs
of Fig.~\ref{figone} the phase space integral in Eq.~(\ref{M34tot}) actually
diverges. The same problem is already encountered for Low-Nussinov Pomeron
exchange in $qQ$ elastic scattering and can be circumvented by replacing the
massless gluon propagator by a regularized version which avoids gluon
propagation over long distances~\cite{lan}. QCD Pomeron models of this kind,
with a dynamically generated gluon mass~\cite{cornwall}, have been found to
give a good description of available data~\cite{natale}. We approximate these
refinements by multiplying the four products ${T^{\dagger}}_{fn}T_{ni}$ in
the unitarity relation (corresponding to the four groups of intermediate
states in Fig.~\ref{figthree}) by factors
\begin{equation}
\frac{q^2}{q^2-m_r^2}\;, \qquad m_r = 300\; {\rm MeV}
\end{equation}
for each internal gluon propagator of momentum $q$. Effectively this
corresponds to a replacement of a massless propagator by a massive one.
Since any complete gauge invariant set of amplitudes is multiplied
by a common factor, gauge invariance is preserved by this procedure.
Finally, the phase space integral of Eq.~(\ref{M34tot}) was performed
numerically by evaluating individual tree level amplitudes with the aid
of the helicity amplitude techniques of Ref.~\cite{HZ}. The numerical
evaluation of this ``loop''-integral is quite slow (of order ten minutes
for one individual phase space point on our Alpha-station) and does not
allow detailed investigations via a Monte Carlo program. However, the code
is quite adequate for a first qualitative study.

\section{Radiation Pattern for Two-Gluon Exchange}

Numerical results for $|M_{34}|^2/s$ are shown in Fig.~\ref{figfour}. The
square of this color singlet exchange amplitude was obtained for particular
kinematic configurations (the same ones as used in Section II) where the two
quarks are held fixed at $p_T=30$~GeV and pseudorapidities $\eta_q=3$ and
$\eta_Q=-3$ as indicated by the arrows. The gluon then is taken at fixed
$p_{Tg}=2\,(15)$~GeV and the gluon rapidity is varied between -8 and +8.
The solid lines depict the gluon angular distribution for the imaginary part
of the two gluon exchange amplitude, $Im(M_{34})$. The distributions follow
neither the ones for $t$-channel photon exchange (dashed lines) nor the ones
for color octet single gluon exchange (dash-dotted lines).
Rather one finds a strong enhancement in the central region, $-3<\eta_g<3$,
which is particularly pronounced for the emission of high transverse momentum
gluons.

This enhancement can be traced to Feynman graphs like the first one in
Fig.~\ref{figfive}, the $A_i^{(5)}D_f^{(3)}$ term. Since the upper quark,
$q$, is
fixed at small scattering angle, this diagram corresponds to $q\to g$ splitting
and subsequent elastic $gQ$ scattering by $t$-channel color singlet
two gluon exchange, {\it i.e.} by the exchange of a Low-Nussinov Pomeron.
There is no reason why this process should be suppressed except for
$\eta_g<\eta_Q$, which corresponds to large scattering angles
in the $gQ$ rest frame. Indeed the fall-off observed in Fig.~\ref{figfour}
at $\eta_g<-3$ confirms this interpretation.

In order to isolate gluon radiation off ``Pomeron exchange'' between the two
quark lines one needs to eliminate the collinear contributions involving
Pomeron exchange between the gluon and the quarks. A full treatment of the
collinear region requires convolution with quark and gluon structure
functions. Because we do not yet have analytic formulas for the
imaginary part or at least a fast Monte Carlo program, we cannot at present
perform the full calculation. However, for our qualitative discussion a crude
approximation will suffice. From the full amplitude $Im(M_{34})$ we subtract
the gluon box diagram $A_i^{(5)}D_f^{(3)}$ of Fig.~\ref{figfive} and the
minimal set of additional graphs required for gauge invariance. We thus define
the Pomeron amplitude in $qQ\rightarrow qQg$ scattering as
\begin{eqnarray}
Im(M_{34}^{\rm Pom}) & = & Im(M_{34})
-\frac{g^5}{16N^2}\frac{1}{32\pi^2}\int d\Omega  \biggl(
4N^2\left[ A_i^{(5)}+A_i^{(3)}-A_i^{(2)} \right]
\left[ D_f^{(1)}+D_f^{(3)} \right]  \biggr) \; .
\label{M34pom}
\end{eqnarray}
The above choice is fixed not by gauge invariance alone, but rather by the
necessity to avoid over-subtraction in all regions of phase space. While
the combination $-D_f^{(2)}+D_f^{(3)}$ for the last factor in
Eq.~(\ref{M34pom}) is gauge invariant also, the concomitant contribution
$A_i^{(5)}D_f^{(2)}$ (see second graph in Fig.~\ref{figfive})
is strongly peaked for back-scattered gluons and thus
would not correspond to Pomeron exchange in $gQ$ scattering. Similarly, the
$s$-channel quark exchange graphs $A_i^{(2)}$ and $A_i^{(3)}$ (see
Fig.~\ref{figone}) are chosen to form a gauge invariant combination with
$A_i^{(5)}$. The additional pieces are numerically small, unlike $A_i^{(1)}$
and $A_i^{(4)}$
terms which receive strong enhancements from the $u$-channel quark propagators
and lead to over-subtraction in the $\eta_g>\eta_q$ and $\eta_g<\eta_Q$
regions, respectively.

The resulting radiation pattern in ``Pomeron exchange'' between the two quark
lines is given in Fig.~\ref{figsix} for three values of the regularizing gluon
mass parameter, $m_r = 0.3$~GeV (solid lines), $m_r = 0.15$~GeV (dotted lines),
and $m_r = 1$~GeV (dash-double dotted lines). The $p_{Tg}=2$~GeV
case shows that soft gluon emission strongly resembles the pattern found for
$t$-channel photon exchange: emission into the gap region between the two
quarks and backwards emission are strongly suppressed.
Soft gluon radiation in Pomeron exchange therefore produces
the pattern expected for $t$-channel color singlet exchange, and as
argued earlier for photon exchange this pattern should be preserved for
multiple soft gluon emission. This strongly suggests that quark scattering by
exchange of a hard Pomeron indeed leads to the formation of rapidity gaps in
dijet events~\cite{bjgap,chehime}.
For hard gluon radiation ($p_{Tg}=15$~GeV in Fig.~\ref{figsix}(b))
backward emission, like in the QCD case, dominates and no suppressed
radiation into the gap region can be expected at higher orders.

Qualitatively, the radiation pattern in Pomeron exchange,
$Im(M_{34}^{\rm Pom})$, can be described as a superposition of a QED-like
component, which dominates at small gluon transverse momenta, and a QCD-like
component, which dominates in the hard region. The transition between the soft
and hard regions is displayed in Fig.~\ref{figseven} by showing
$|M_{34}|^2/s$ as a function of $p_{Tg}$ for the same quark configuration as
in Fig.~\ref{figsix}. In the  back-scattering region (see
Fig.~\ref{figseven}(a) the emission of soft gluons is severely suppressed
compared to the tree level QCD case. In the forward region (see
Fig.~\ref{figseven}(b) soft gluon emission in ``Pomeron exchange'' essentially
follows the shape expected for photon exchange. For large gluon transverse
momenta, however, the QED-like
component of $Im(M_{34}^{\rm Pom})$ is severely suppressed.

The more QCD-like
behavior at large $p_{Tg}$ may be understood by noting that the phase space
integral for color singlet two-gluon exchange in Eq.~(\ref{M34tot}) is
dominated by the region of small $|q^2|$ for one of the two gluons. A hard
radiated gluon is insensitive to the color screening by this soft gluon and
effectively only the color charge of the second, hard $t$-channel gluon is
seen, leading to a more QCD-like radiation pattern.

This interpretation is confirmed in Figs.~\ref{figsix} and \ref{figseven}
by the curves for
different values of the regularizing gluon mass. Larger values of $m_r$
eliminate the soft gluon exchange region where the
emitted gluon can resolve the color charges of the two $t$-channel gluons.
Thus the color singlet structure of $t$-channel two gluon exchange gains
importance and the radiation pattern becomes more QED-like.

Figs.~\ref{figsix} and \ref{figseven} also exhibit the dependence of our
results
on the regularization of the gluon propagator.  In the regions of strong
destructive interference in $Im(M_{34})$, which are visible in
Fig.~\ref{figsix}, the quantitative results strongly depend on the
details of soft gluon exchange and are thus non-perturbative in nature. The
qualitative features, however, suppression of backwards soft gluon
emission in color singlet two gluon exchange events and a more
QCD-like behavior in the hard region, are insensitive to these uncertainties.
In particular, the variation with $m_r$ is modest in the most important
region, forward gluon emission at small $p_{Tg}$. Hence, our main
finding, the QED like radiation pattern of soft gluons in Pomeron
exchange, is unaffected.

\section{Conclusions }

We have analyzed the different properties of $t$-channel color singlet $vs.$
color octet exchange processes in $qQ$ scattering at next to leading order,
namely including gluon bremsstrahlung. The radiation pattern of emitted
gluons is intimately connected, after hadronization, to the angular
distribution of produced hadrons and hence directly bears on the question
whether rapidity gaps are to be expected in $t$-channel color singlet
exchange processes. Essentially, we arrive at an affirmative answer, also
for the case of color singlet two gluon exchange.

While this answer was expected by many in the field, our calculations reveal
a rich internal structure of the Pomeron. In addition, our demonstration had
to deal with a number of problems which do not yet appear at the level of
$2\to 2$ processes. While the definition of $t$-channel
color singlet exchange is
obvious in the case of $qQ\to qQ$ scattering, the emission of an extra
gluon complicates the picture. The color tensors of the $qQ\to qQg$ process
suggest identification of color singlet exchange as those terms where the
color of at least one of the incident quarks remains unchanged. In the QCD
case (single gluon exchange) this also occurs since the color of the
exchanged gluon can be compensated for by emission of a gluon of the same
color. The crucial difference emerges when considering the angular
distribution of emitted gluons: $t$-channel photon exchange leads to
forward gluon emission and hence preserves the spatial distribution of
the color charges which are present at lowest order. By contrast gluons
are radiated in the backwards direction in the QCD color singlet term, and
multiple emission will wash out the pattern produced by the first emitted
gluon.

Color singlet and color octet exchange terms do not interfere because they
are orthogonal in color. Because of their different behaviour under multiple
gluon emission any interference between a QED- and a QCD-like ``color singlet
pattern'' will be severely suppressed at higher order. The square of the
QED-like component in the color singlet exchange amplitude can hence be
treated as the probability density for producing a hard scattering event
which may evolve into a rapidity gap signature. For technical reasons
(numerical speed) we cannot yet calculate rapidity gap cross sections,
rather we are limited to an analysis at the amplitude square level when
considering the $t$-channel exchange of two gluons.

For $t$-channel two gluon exchange we have calculated the imaginary part
of the color singlet exchange amplitude, $Im(M_{34})$, and we have analyzed
the emission probability for the extra gluon when the quarks are scattered
by small angles only. At first sight gluon emission is most likely in the
angular region between the two quarks. However, this pattern is solely due
to $q\to g$ splitting and subsequent $gQ$ scattering via the exchange of a
Low-Nussinov Pomeron. Only after subtracting this contribution does the final
pattern emerge for gluon emission in $qQ$ scattering via Pomeron exchange.

For high transverse momentum of the emitted gluon (of order of the quark
momenta) the radiation pattern is quite similar to the one obtained for
single gluon exchange.  Hard
emitted gluons have too short a wavelength to see the screening of the
color charge of the harder exchanged gluon by the second, typically very
soft, exchanged gluon. The Low-Nussinov Pomeron thus reveals itself as an
extended object. 
Hard gluon emission is able to resolve the internal color structure of the
Pomeron.

As the transverse momentum of the emitted gluon is decreased, a qualitative
transition occurs. The gluon radiation can no longer resolve this internal
color structure and hence the Pomeron appears as a color singlet object.
As a result the
emission of a soft gluon ($p_{Tg}<<p_{Tq}$) follows a pattern very similar
to the one observed for $t$-channel photon exchange. This pattern is
expected to lead to the formation of rapidity gaps. Since the overall gluon
emission rate is dominated by the soft region, we conclude that two gluon
color singlet exchange in dijet events may indeed lead to the formation of
rapidity gap events as observed at the Tevatron~\cite{brandt}.

\newpage
{\bf Acknowledgements}
We thank J.~D.~Bjorken, S.~Ellis, F.~Halzen, and G.~Sterman for helpful
discussions.
This research was supported in part by the University of Wisconsin Research
Committee with funds granted by the Wisconsin Alumni Research Foundation,
by the U.~S.~Department of Energy under contract No.~DE-AC02-76ER00881,
and by the Texas National Research Laboratory Commission under Grants
No.~RGFY9273 and FCFY9212.

\figure{Feynman graphs for the process $qQ\to qQg$ via $t$-channel gluon
exchange.      \label{figone} }

\figure{Rapidity distribution of emitted gluons in $uc\to ucg$ scattering.
The final state quarks are kept fixed at forward rapidities of $\eta_q =
\pm 3$ (indicated by the arrows) and transverse momenta $p_{Tq}=30$~GeV while
the gluon rapidity is varied between -8 and 8 at fixed transverse momentum
$p_{Tg} = 2$~GeV. In part a) results are shown for the sum over all color
structures for single gluon exchange (dash-double-dotted line) and for
$t$-channel photon exchange (dashed line). In addition the QCD color singlet
exchange contribution, as defined by the $M_{12}$ and $M_{34}$ terms in
Eq.~(\ref{sumM2}), is shown (dash-dotted line). The
$M_{34}$ terms alone, in part b), demonstrate the difference between the QED
and the QCD color singlet exchange terms.         \label{QEDQCDtree} }

\figure{Feynman graphs contributing to the imaginary part of the $qQ\to qQg$
amplitude at order $g_{s}^{5}$. The four groups correspond to $qQ$
intermediate states (a) and three parton intermediate states with subsequent
$qQ$ scattering (b), $qg$ scattering (c), and $Qg$ scattering (d). The dashed
vertical line shows where to make the cut in order to obtain the imaginary
part. The corresponding product of tree level $2\to 2$ and $2\to 3$
amplitudes is given below each graph. The crosses on the propagators of
the $2\to 3$ tree level sub-amplitudes indicate the other positions where
the external gluon propagator needs to be attached.       \label{figthree} }

\figure{Rapidity distribution of emitted gluons in $uc\to ucg$ scattering.
Results are shown for the color singlet contribution as seen by the charm
quark ($12|M_{34}|^2/s$).
The final state quarks are kept fixed at forward rapidities of $\eta_q =
\pm 3$ and transverse momenta $p_{Tq}=30$~GeV while the gluon rapidity
is varied between -8 and 8. Results are shown for the case of a)
a ``soft'' gluon of $p_{Tg}=2$~GeV and b) a hard gluon of $p_{Tg}=15$~GeV.
The solid lines give the results for the full imaginary part of the color
singlet exchange amplitude $M_{34}$. For comparison the tree level results for
$12|M_{34}|^2/s$ are shown for gluon exchange (dash-dotted lines) and photon
exchange (dashed lines).         \label{figfour} }

\figure{Two of the Feynman graphs corresponding to $q\rightarrow g$ splitting
and subsequent $gQ$ scattering.   \label{figfive} }

\figure{Rapidity distribution of emitted gluons in $uc\to ucg$ scattering.
Results are shown for the color singlet contribution as seen by the charm
quark ($12|M_{34}|^2/s$) after subtraction of the ``Pomeron'' contribution
to $u\to g$ splitting and subsequent $cg$ scattering (see text). The dotted,
solid, and dash-double-dotted lines show the gluon radiation pattern for three
different values of the regularizing gluon mass, $m_r$. The phase space
configurations are the same as in Fig.~\ref{figfour}. For comparison tree
level results for $12|M_{34}|^2/s$ are shown for gluon exchange
(dash-dotted lines) and photon exchange (dashed lines).
          \label{figsix} }

\figure{Dependence of the color singlet exchange amplitude, $12|M_{34}|^2/s$,
on the gluon transverse momentum $p_{Tg}$ at two values of the gluon scattering
angle, a) $\eta_g=-5$ and b) $\eta_g=+5$. The final state quarks are fixed
at the same momenta as in Fig.~\ref{figfour}. Results are shown for three
values of the regularizing gluon mass, $m_r=0.15$~GeV (dotted line),
$m_r=0.3$~GeV (solid line), and $m_r=1$~GeV (dash-double-dotted line). For
comparison, the corresponding QCD and QED distributions are also shown.
         \label{figseven} }

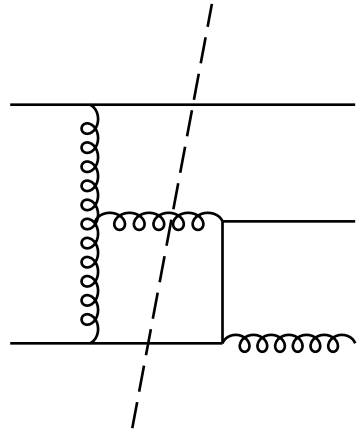
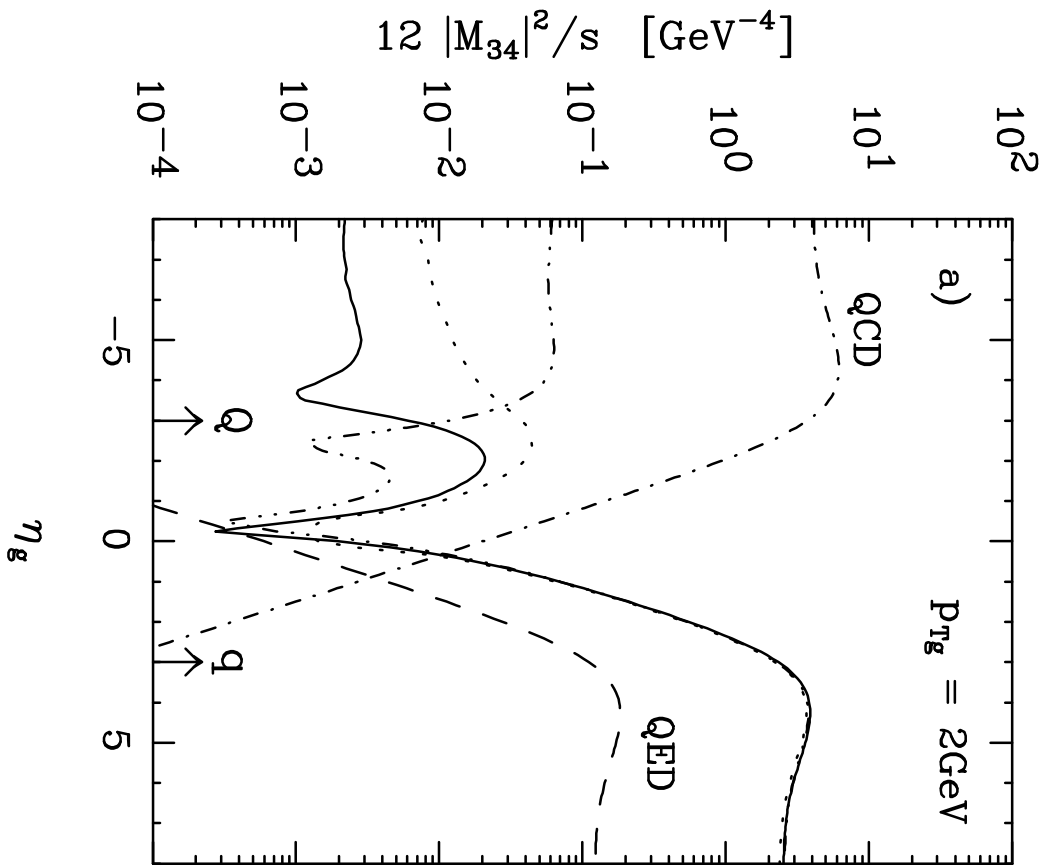
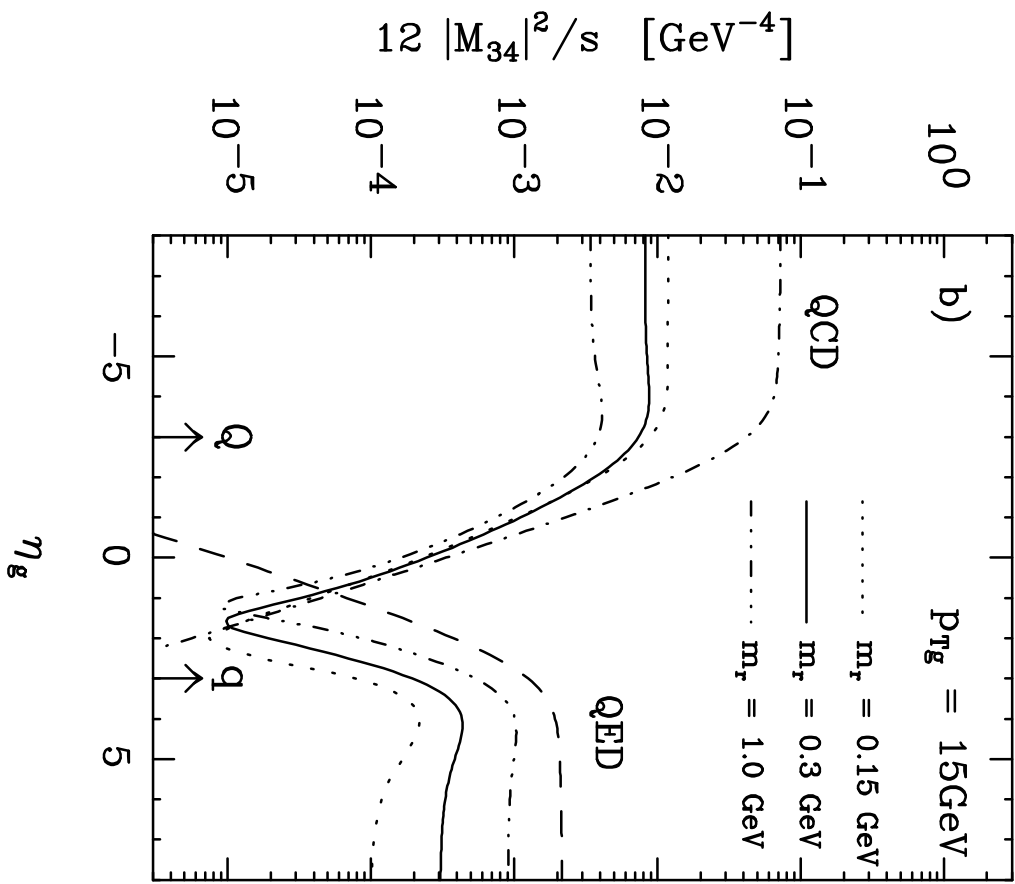
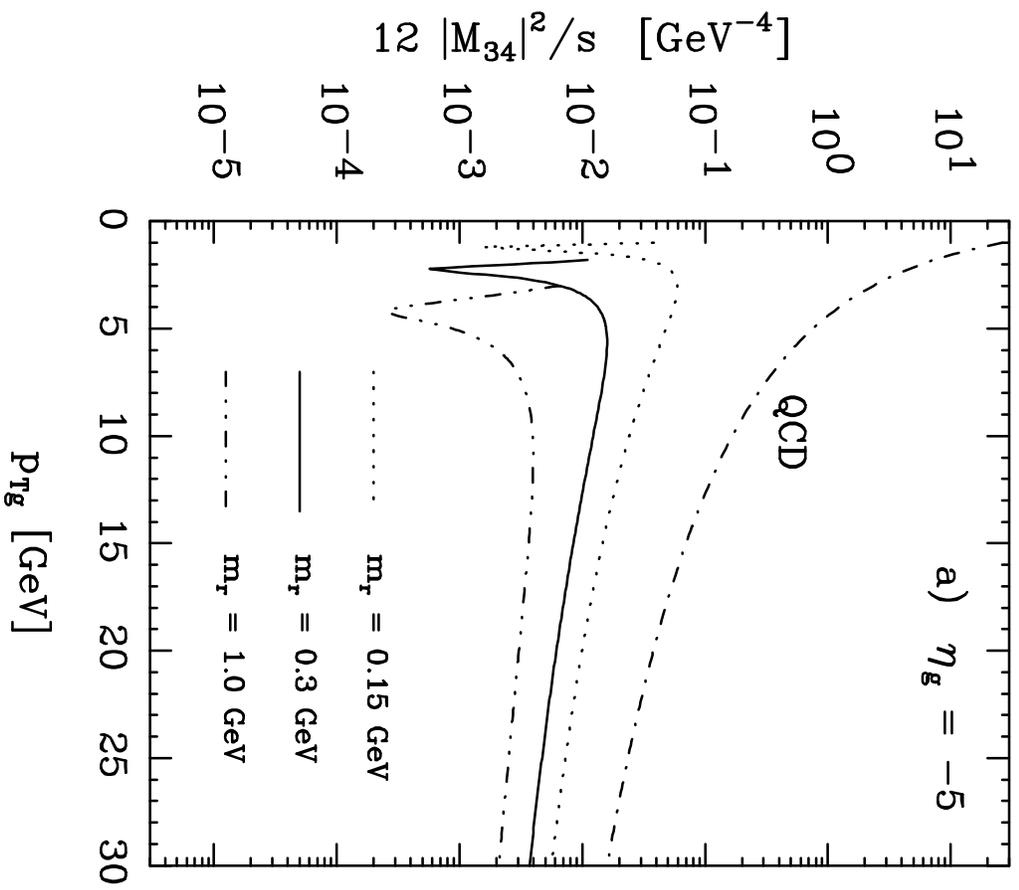
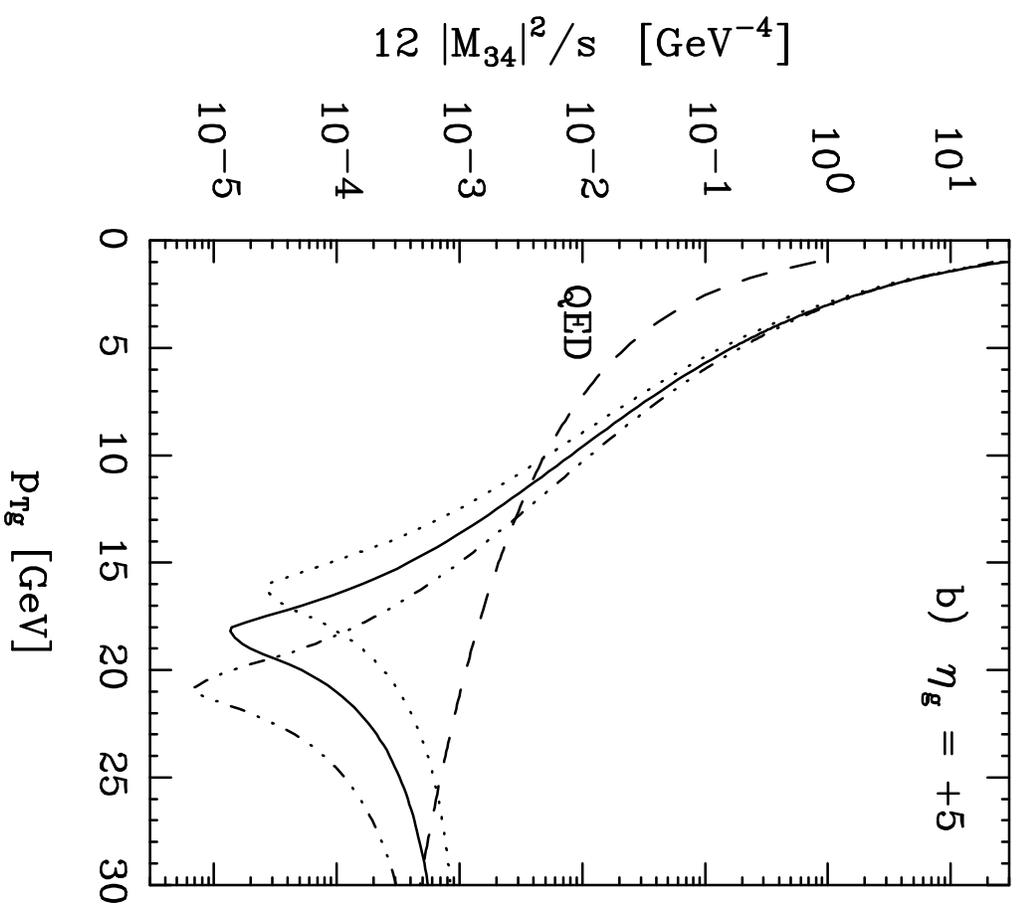
\end{document}